 \documentclass{cambridge6A}
 \usepackage{natbib}
  \usepackage{rotating}
  \usepackage{floatpag}
  \rotfloatpagestyle{empty}
  \usepackage{amsthm}
  \usepackage{graphicx}

\usepackage{url}

\def\lsim{\lower.5ex\hbox{$\; \buildrel < \over \sim \;$}}
\def\gsim{\lower.5ex\hbox{$\; \buildrel > \over \sim \;$}}

\def\mcrx{{\it Sunrise}}

\begin{document}

\title{Cosmological Structure Formation}
\author{Joel R. Primack\\[3\baselineskip]
\affiliation{Physics Department, University of California, Santa Cruz, CA 95064 USA.}}

  \maketitle
\mainmatter
\chapter{Cosmological Structure Formation}

\section{Introduction}
Cosmology has finally become a mature science during the past decade, with predictions now routinely confirmed by observations.  The modern cosmological theory is known as $\Lambda$CDM -- CDM for Cold Dark Matter, particles that moved sluggishly in the early universe and thereby preserved fluctuations down to small scales \citep[][see Fig. 1]{BFPR84}, and $\Lambda$ for the cosmological constant \citep[e.g.,][]{LLPR91}.  A wide variety of large-scale astronomical observations -- including the Cosmic Microwave Background (CMB),  measurements of baryon acoustic oscillations (BAO), gravitational lensing, the large-scale distribution of galaxies, and the properties of galaxy clusters -- agree very well with the predictions of the $\Lambda$CDM cosmology.  

Like the standard model of particle physics, the $\Lambda$CDM standard cosmological model requires the determination of a number of relevant cosmological parameters, and the theory does not attempt to explain why they have the measured values -- or to explain the fundamental nature of the dark matter and dark energy.  These remain challenges for the future.  But the good news is that the key cosmological parameters are now all determined with unprecedented accuracy, and the six-parameter $\Lambda$CDM theory provides a very good match to all the observational data including the 2015 Planck temperature and polarization data \citep{Planck15-XIII}.  Within uncertainties less than 1\%, the Universe has critical cosmic density -- i.e., $\Omega_{\rm total} = 1.00$ and the Universe is Euclidean (or ``flat") on large scales.  The expansion rate of the Universe is measured by the Hubble parameter $h = 0.6774\pm0.0046$, and $\Omega_{\rm matter} = 0.3089\pm0.0062$; this leads to the age of the Universe $t_0=13.799\pm0.021$ Gyr.  The power spectrum normalization parameter is $\sigma_8=0.816\pm0.009$, and the primordial fluctuations are consistent with a purely adiabatic spectrum of fluctuations with a spectral tilt $n_s = 0.968 \pm 0.006$, as predicted by single-field inflationary models \citep{Planck15-XIII}.  The same cosmological parameters that are such a good match to the CMB observations also predict the observed distribution of density fluctuations from small scales probed by the Lyman alpha forest\footnote{The Ly$\alpha$ forest is the many absorption lines in quasar spectra due to clouds of neutral hydrogen along the line of sight to the quasar.} to the entire horizon, as shown in Fig. 2.  The near-power-law galaxy-galaxy correlation function at low redshifts is now known to be a cosmic coincidence \citep{Watson11}.  I was personally particularly impressed that the evolution of the galaxy-galaxy correlations with redshift predicted by $\Lambda$CDM \citep{Kravtsov04} turned out to be in excellent agreement with the subsequent observations \citep[e.g.,][]{ConroyWechslerKravtsov06}.

For non-astronomers, there should be a more friendly name than $\Lambda$CDM for the standard modern cosmology.  Since about 95\% of the cosmic density is dark energy (either a cosmological constant with $\Omega_\Lambda =0.69$ or some dynamical field that plays a similar cosmic role) and cold dark matter with $\Omega_{\rm CDM} = 0.26$, I recommend the simple name ``Double Dark Theory" for the modern cosmological standard model \citep{PrimackAbrams,AbramsPrimack}.  The contribution of ordinary baryonic matter is only $\Omega_{\rm b} = 0.05$.  Only about 10\% of the baryonic matter is in the form of stars or gas clouds that emit electromagnetic radiation, and the contribution of what astronomers call ``metals" -- chemical elements heavier than helium -- to the cosmic density is only $\Omega_{\rm metals} \approx 0.0005$, most of which is in white dwarfs and neutron stars \citep{FukugitaPeebles}.  The contribution of neutrino mass to the cosmic density is $0.002 \le \Omega_\nu \le 0.005$, far greater than $\Omega_{\rm metals}$.  Thus our bodies and our planet are made of the rarest form of matter in the universe: elements forged in stars and stellar explosions.

\begin{figure}[t!]
\includegraphics[width=\linewidth]{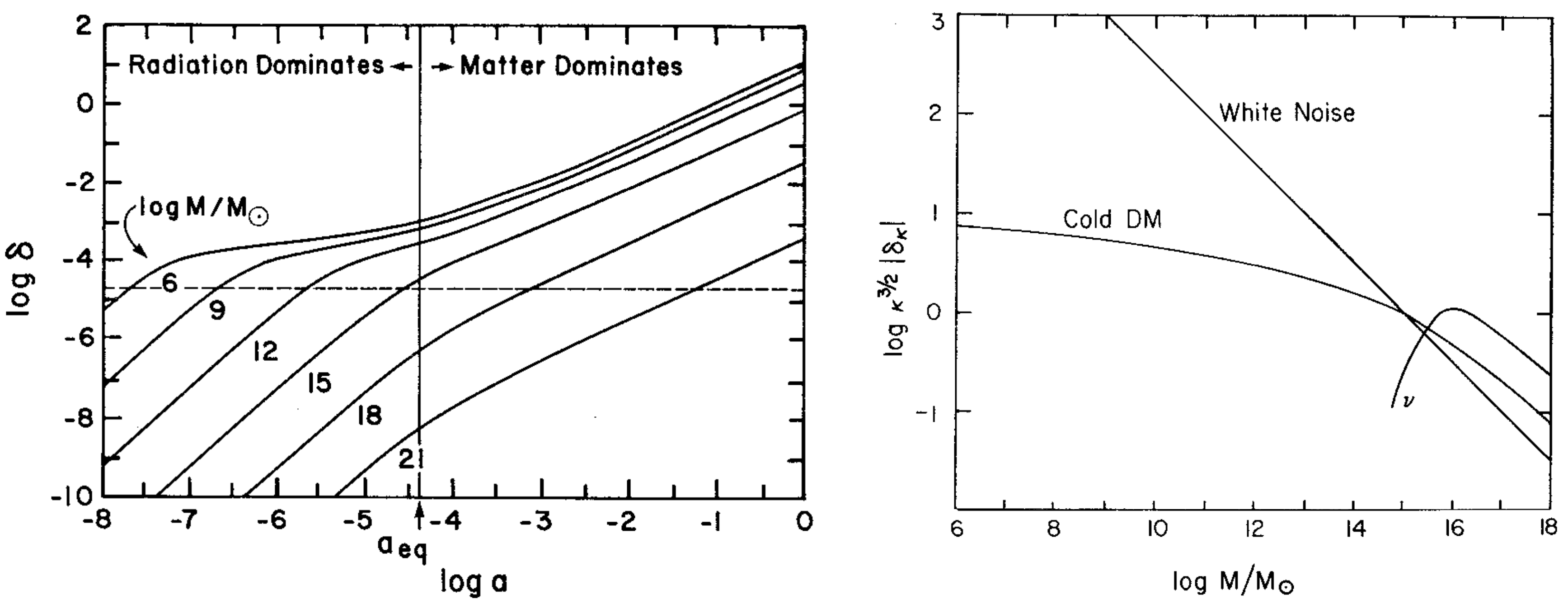}
\caption{The origin of the CDM spectrum of density fluctuations.  {\bf Left panel}: Fluctuations corresponding to mass scales $10^6 M_\odot$, $10^9 M_\odot$, etc., grow proportionally to the square of scale factor $a$ when they are outside the horizon, and when they enter the horizon (cross the horizontal dashed line) the growth of the fluctuation amplitude $\delta$ is much slower if they enter when the Universe is radiation dominated (i.e., $a < a_{\rm eq}$).  Fluctuations on mass scales $> 10^{15} M_\odot$ enter the horizon after it becomes matter dominated, so their growth is proportional to scale factor $a$; that explains the larger separation between amplitudes for such higher-mass fluctuations.  (From a 1983 conference presentation \citet{PrimackBlum83}, reprinted in \citet{Varenna84}.) {\bf Right panel}: The resulting CDM fluctuation spectrum ($\kappa^{3/2} |\delta_\kappa| = \Delta M/M$) is contrasted with a $\delta \propto \kappa^0$ white noise (Poisson) spectrum with the same power at all wavelengths (where wave number $\kappa$ is as usual related to wavelength $\lambda$ by $\kappa = 2\pi/\lambda$), and with the hot dark matter spectrum if the dark matter were light neutrinos ($\nu$) which is cut off on galaxy mass scales by free-steaming.  (From \cite{BFPR84}.  This calculation assumed that the primordial fluctuations are scale-invariant (Zel'dovich) and that $\Omega_{\rm matter} = 1$ and Hubble parameter $h=1$.) 
}
\label{fig:1}
\end{figure}

\begin{figure}[t!]
\includegraphics[width=\linewidth]{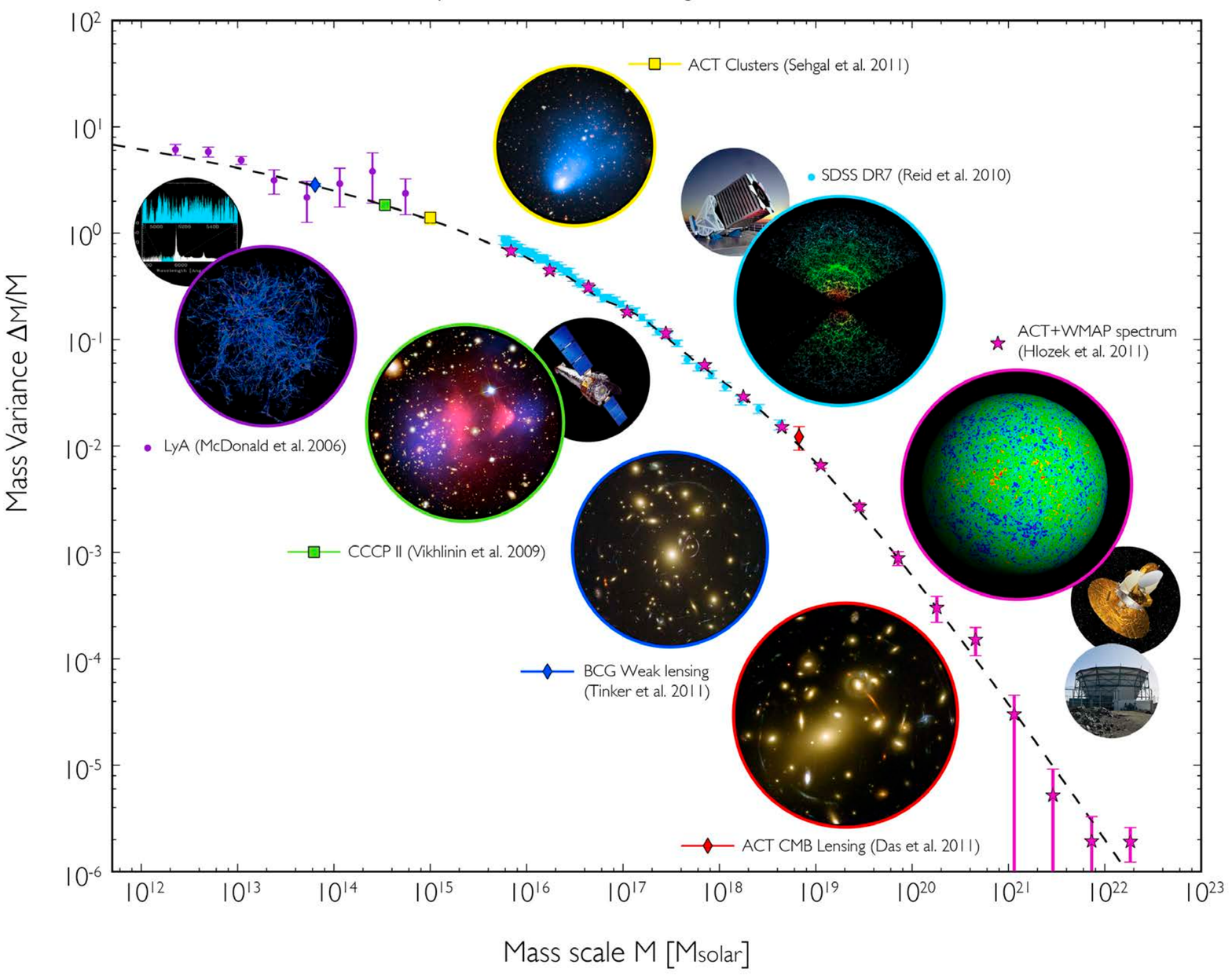}
\caption{The r.m.s.~mass variance $\Delta M/M$ predicted by $\Lambda$CDM compared with observations, from CMB and the Atacama Cosmology Telescope (ACT) on large scales, brightest cluster galaxy weak lensing, clusters, the SDSS galaxy distribution, to the Lyman alpha forest on small scales.  This figure highlights the consistency of power spectrum measurements by an array of cosmological probes over a large range of scales.  (Redrawn from Fig. 5 in \citet{Hlozek}, which gives the sources of the data.)}
\label{fig:2}
\end{figure}

Potential challenges to $\Lambda$CDM on large scales come from the tails of the predicted distribution functions, such as CMB cold spots and massive clusters at high redshifts. However, the existing observations appear to be consistent thus far with predictions of standard $\Lambda$CDM with standard primordial power spectra; non-Gaussian initial conditions are not required \citep{Planck15-XVII}.  Larger surveys now underway may provide more stringent tests.

\section{Large-Scale Structure}

Large, high-resolution simulations permit detailed predictions of the distribution and properties of galaxies and clusters.  From 2005-2010, the benchmark simulations were Millennium-I \citep{Springel05} and Millennium-II \citep{Boylan-Kolchin09}, which have been the basis for more than 400 papers.  However, these simulations used first-year Wilkinson Microwave Anisotropy Probe (WMAP) cosmological parameters, including $\sigma_8=0.90$, that are now in serious disagreement with the latest observations.  Improved cosmological parameters, simulation codes, and computer power have permitted more accurate simulations \citep{Kuhlen12,DarkSky14} including Bolshoi \citep{Klypin+11} and BigBolshoi/MultiDark \citep{Prada12,MultiDark}, which have recently been rerun using the Planck cosmological parameters \citep{Klypin14}.\footnote{The web address for the MultiDark simulation data center is \url{http://www.cosmosim.org/cms/simulations/multidark-project/}}  

Dark matter halos can be characterized in a number of ways.  A common one is by mass, but the mass attributed to a halo depends on a number of factors including how the outer edge of the halo is defined; popular choices include the spherical radius within which the average density is either 200 times critical density or the virial density (which depend on redshift).  Properties of all the halos in many stored time steps of both the Bolshoi and BigBolshoi/MultiDark simulations are available on the web in the MultiDark database.$^2$  For many purposes it is more useful to characterize halos by their maximum circular velocity 
$V_{\rm max}$, which is defined as the maximum value of $[G M(<r)/r]^{1/2}$, where $G$ is Newton's constant and $M(<r)$ is the mass enclosed within radius $r$.  The reason this is useful is that $V_{\rm max}$ is reached at a relatively low radius $r_{\rm max}$, closer to the central region of a halo where stars or gas can be used to trace the velocity of the halo, while most of the halo mass is at larger radii.  Moreover, the measured internal velocity of a galaxy (line of sight velocity dispersion for early-type galaxies and rotation velocity for late-type galaxies) is closely related to its luminosity according to the Faber-Jackson and Tully-Fisher relations.  In addition, after a subhalo has been accreted by a larger halo, tidal stripping of its outer parts can drastically reduce the halo mass but typically decreases $V_{\rm max}$ much less.  (Since the stellar content of a subhalo is thought to be determined before it was accreted, some authors define $V_{\rm max}$ to be the peak value at any redshift for the main progenitor of a halo.)  Because of the observational connection between larger halo internal velocity and brighter galaxy luminosity, a common simple method of assigning galaxies to dark matter halos and subhalos is to rank order the galaxies by luminosity and the halos by $V_{\rm max}$, and then match them such that the number densities are comparable \citep{Kravtsov04,Tasitsiomi,ConroyWechslerKravtsov06}.  This is called ``halo abundance matching" or (more modestly) ``sub-halo abundance matching" (SHAM) \citep{Reddick14}.  Halo abundance matching using the Bolshoi simulation predicts galaxy-galaxy correlations (which are essentially counts of the numbers of pairs of galaxies at different separation distances) that are in good agreement with the Sloan Digital Sky Survey (SDSS) observations \citep{Trujillo-Gomez11, Reddick13}. 

Abundance matching with the Bolshoi simulation also predicts galaxy velocity-mass scaling relations consistent with observations \citep{Trujillo-Gomez11}, and a galaxy velocity function in good agreement with observations for maximum circular velocities $V_{\rm max} \gsim 100$ km/s, but higher than the HI Parkes All Sky Survey (HIPASS) and the Arecibo Legacy Fast ALFA (ALFALFA) Survey radio observations \citep{Zwaan10,Papastergis11} by about a factor of 2 at 80 km/s and a factor of 10 at 50 km/s.  This either means that these radio surveys are increasingly incomplete at lower velocities, or else $\Lambda$CDM is in trouble because it predicts far more small-$V_{\rm max}$ halos than there are observed low-$V$ galaxies.  A deeper optical survey out to 10 Mpc found no disagreement between $V_{\rm max}$ predictions and observations for $V_{\rm max} \ge 60$ km/s, and only a factor of 2 excess of halos compared to galaxies at 40 km/s \citep{KlypinKarachentsev}.  This may indicate that there is no serious inconsistency with theory, since for $V \lsim 30$ km/s reionization and feedback can plausibly explain why there are fewer observed galaxies than dark matter halos \citep{BKW2000,Somerville02,Benson03,Kravtsov10,WadepuhlSpringel11, Sawala12}, and also the observed scaling of metallicity with galaxy mass \citep{DekelWoo03,Woo08,Kirby11}.

The radial dark matter density distribution in halos can be approximately fit by the simple formula 
$\rho_{\rm NFW} = 4\rho_s x^{-1}(1+x)^{-2}$, where $x\equiv r/r_s$ \citep{NFW}, and the ``concentration" of a dark matter halo is defined as $C=R_{\rm vir}/R_s$ where $R_{\rm vir}$ is the virial radius of the halo. When we first understood that dark matter halos form with relatively low concentration $C \sim 4$ and evolve to higher concentration, we suggested that ``red" galaxies that shine mostly by the light of red giant stars because they have stopped forming stars should be found in high-concentration halos while ``blue" galaxies that are still forming stars should be found in younger low-concentration halos \citep{Bullock01}.  This idea was recently rediscovered by Hearin and Watson, who used the Bolshoi simulation to show that this leads to remarkably accurate predictions for the correlation functions of red and blue galaxies \citep{AgeMatching13,AgeMatching14}.

The Milky Way has two rather bright satellite galaxies, the Large and Small Magellanic Clouds.  It is possible using sub-halo abundance matching with the Bolshoi simulation to determine the number of Milky-Way-mass dark matter halos that have subhalos with high enough circular velocity to host such satellites.  It turns out that about 55\% have no such subhalos, about 28\% have one, about 11\% have two, and so on \citep{Busha11a}.  Remarkably, these predictions are in excellent agreement with an analysis of observations by the Sloan Digital Sky Survey (SDSS) \citep{Liu11}.  
The distribution of the relative velocities of central and bright satellite galaxies from SDSS spectroscopic observations is also in very good agreement with the predictions of the Millennium-II simulation \citep{Tollerud11}, and the Milky Way's lower-luminosity satellite population is not unusual \citep{StrigariWechsler12}.  Considered in a cosmological context, the Magellanic clouds are likely to have been accreted within about the last Gyr \citep{Besla12}, and the Milky Way halo mass is $1.2^{+0.7}_{-0.4}$(stat.)$\pm0.3$(sys.)$\times10^{12} M_\odot$ \citep{Busha11b}.

\begin{figure}[t!]
\includegraphics[width=\linewidth]{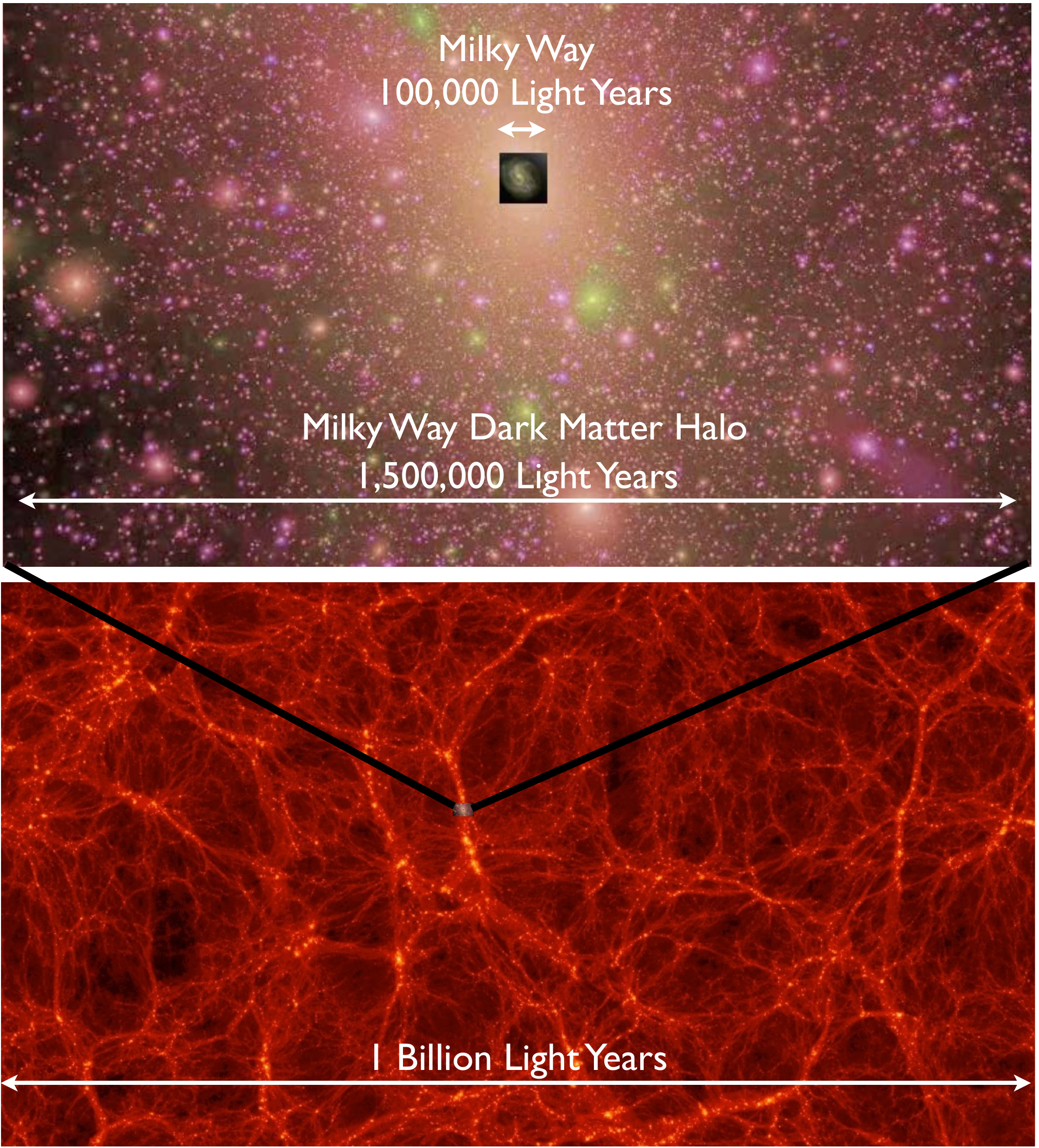}
\caption{The stellar disk of a large spiral galaxy like the Milky Way is about 100,000 light years across, which is tiny compared with the dark matter halo of such a galaxy \cite[from the Aquarius dark matter simulation][]{Aquarius}, and even much smaller compared with the large-scale cosmic web \citep[from the Bolshoi simulation][]{Klypin+11}.
}
\label{fig:3}
\end{figure}

\begin{figure}[b!]
\includegraphics[width=\linewidth]{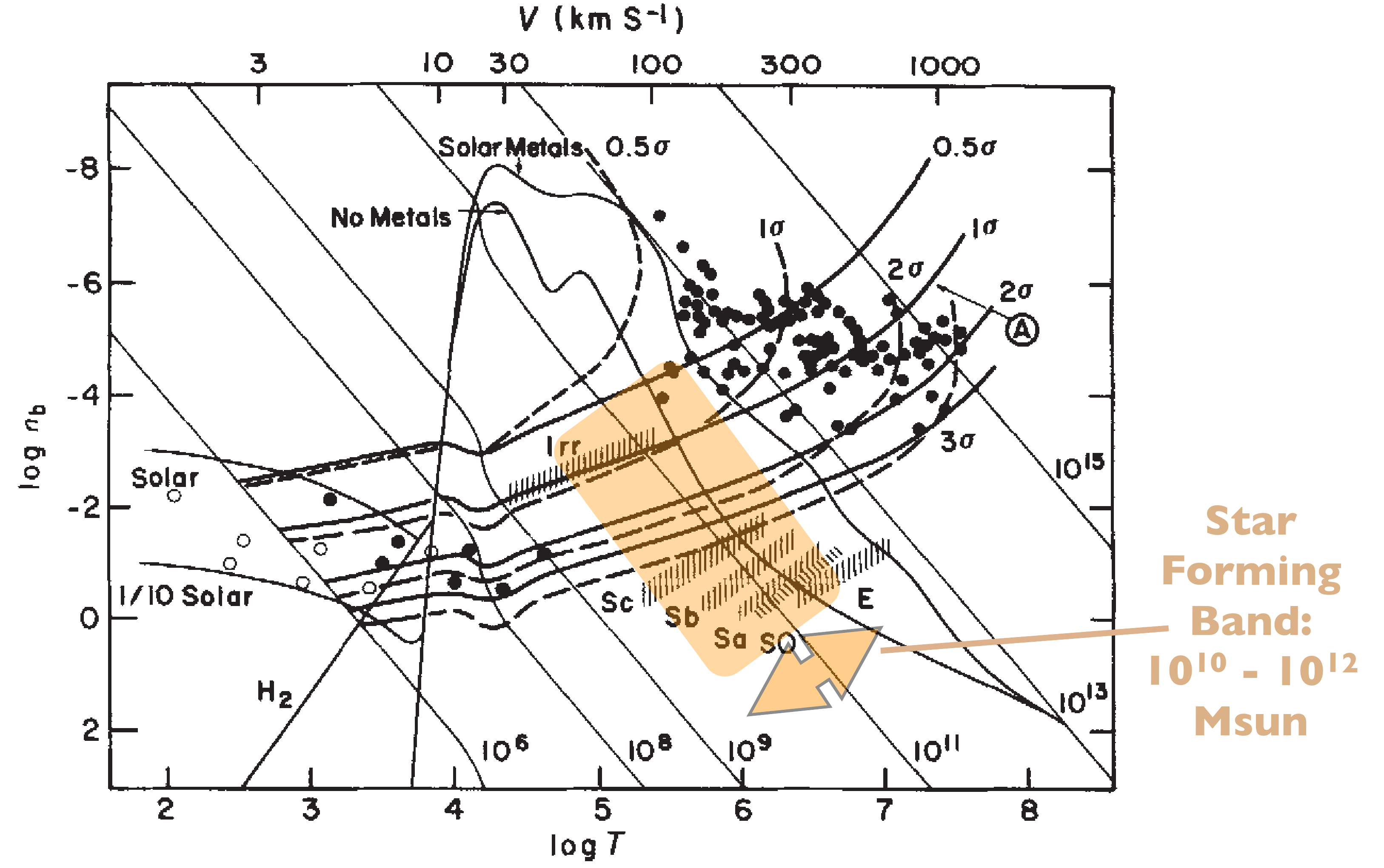}
\caption{The Star-Forming Band on a diagram of baryon density $n_b$ versus the three-dimensional r.m.s.~velocity dispersion $V$ and virial temperature $T$ for structures of various sizes in the universe, where $T=\mu V^2/3k$, $\mu$ is mean molecular weight ($\approx 0.6$ for ionized primordial H + He) and $k$ is Boltzmann's constant. Below the No Metals and Solar Metals cooling curves, the cooling timescale is more rapid than the gravitational timescale. Dots are groups and clusters.
Diagonal lines show the halo masses in units of $M_\odot$.  (This is Fig.~3 in Blumenthal et al., 1984, with the Star-Forming Band added.)}
\label{fig:4}
\end{figure}

\section{Galaxy Formation}

At early times, for example the CMB epoch about 400,000 years after the big bang, or on very large scales at later times, linear calculations starting from the $\Lambda$CDM fluctuation spectrum allow accurate predictions.  But on scales where structure forms, the fluctuations have grown large enough that they are strongly nonlinear, and we must resort to simulations.  The basic idea is that regions that start out with slightly higher than average density expand a little more slowly than average because of gravity, and regions that start out with slightly lower density expand a little faster.  Nonlinear structure forms by the process known by the somewhat misleading name ``gravitational  collapse" -- misleading because what really happens is that when positive fluctuations have grown sufficiently that they are about twice as dense as typical regions their size, they stop expanding while the surrounding universe keeps expanding around them.  The result is that regions that collapse earlier are denser than those that collapse later; thus galaxy dark matter halos are denser than cluster halos.  The visible galaxies form because the ordinary baryonic matter can radiate away its kinetic energy and fall toward the centers of the dark matter halos; when the ordinary matter becomes dense enough it forms stars.  Thus visible galaxies are much smaller than their host dark matter halos, which in turn are much smaller than the large scale structure of the cosmic web, as shown in Fig. 3.

Astronomical observations represent snapshots of moments long ago when the light we now observe left distant astronomical objects.  It is the role of astrophysical theory to produce movies -- both metaphorical and actual -- that link these snapshots together into a coherent physical picture.  To predict cosmological large-scale structures, it has been sufficient to treat all the mass as dark matter in order to calculate the growth of structure and dark matter halo properties.  But hydrodynamic simulations -- i.e., including baryonic matter -- are necessary to treat the formation and evolution of galaxies.

An old criticism of $\Lambda$CDM has been that the order of cosmogony is wrong: halos grow from small to large by accretion in a hierarchical formation theory like $\Lambda$CDM, but the oldest stellar populations are found in the most massive galaxies -- suggesting that these massive galaxies form earliest, a phenomenon known as ``downsizing" \citep{Cowie+96}.  The key to explaining the downsizing phenomenon is the realization that star formation is most efficient in dark matter halos with masses in the band between about $10^{10}$ and 
$10^{12} M_\odot$ \citep[Fig. 1 bottom in][]{BehrooziWechslerConroy13L}.  This goes back at least as far as the original Cold Dark Matter paper \citep{BFPR84}: see Fig.~4.  A dark matter halo that has the total mass of a cluster of galaxies today will have crossed this star-forming mass band at an early epoch, and it will therefore contain galaxies whose stars formed early.  These galaxies will be red and dead today.  A less massive dark matter halo that is now entering the star-forming band today will just be forming significant numbers of stars, and it will be blue today.  The details of the origin of the star-forming band are still being worked out.  Back in 1984, we argued that cooling would be inefficient for masses greater than about $10^{12} M_\odot$ because the density would be too low, and inefficient for masses less than about $10^8 M_\odot$ because the gas would not be heated enough by falling into these small potential wells.  Now we know that reionization, supernovae \citep{DekelSilk86}, and other energy input additionally impedes star formation for halo masses below about $10^{10} M_\odot$, and feedback from active galactic nuclei (AGN) additionally impedes star formation for halo masses above about $10^{12} M_\odot$.

Early simulations of disk galaxy formation found that the stellar disks had much lower rotation velocities than observed galaxies \citep{NavarroSteinmetz00}.  This problem seemed so serious that it became known as the ``angular momentum catastrophe.'' A major cause of this was excessive cooling of the gas in small halos before they merged to form larger galaxies \citep{MallerDekel02}.  Simulations with better resolution and more physical treatment of feedback from star formation appear to resolve this problem.  In particular, the Eris cosmological simulation \citep{Guedes11} produced a very realistic spiral galaxy, as have many simulations since then.  \citet{SomervilleDave14} is an excellent recent review of progress in understanding galaxy formation.  In the following I summarize some of the latest developments.  There are now two leading approaches to simulating galaxies:  
\begin{itemize}
\item{Low resolution, $\sim 1$ kiloparsecs.}  The {\bf advantages} of this approach are that it is possible to simulate many galaxies and study galaxy populations and their interactions with the circumgalactic and intergalactic media, but the {\bf disadvantages} are that we learn relatively little about how galaxies themselves form and evolve at high redshifts.  The prime examples of this approach now are the {\it Illustris} \citep{Illustris} and EAGLE \citep{EAGLE} simulations.  Like semi-analytic models of galaxy formation \citep[reviewed in][]{Benson10}, these projects adjusted the parameters governing star-formation and feedback processes in order to reproduce key properties of galaxies at the present epoch, redshift $z=0$.  The {\it Illustris} simulation in a volume $(106.5 {\rm Mpc})^3$ forms $\sim 40,000$ galaxies at the present epoch with a reasonable mix of elliptical and spiral galaxies that  have realistic appearances \citep{SnyderIllustris15}, obey observed scaling relations, have the observed numbers of galaxies as a function of their luminosity, and were formed with the observed cosmic star formation rate \citep{Illustris14}.   It forms massive compact galaxies by redshift $z=2$ via central starbursts in major mergers of gas-rich galaxies or else by assembly at very early times \citep{Wellons}.  Remarkably, the {\it Illustris} simulation also predicts a population of damped Lyman $\alpha$ absorbers (DLAs, small-galaxy-size clouds of neutral hydrogen) that agrees with some of the key observational properties of DLAs \citep{Bird14,Bird15}.  The EAGLE simulation in volumes up to $(100 {\rm Mpc})^3$ reproduces the observed galaxy mass function from $10^8$ to $10^{11} M_\odot$ at a level of agreement close to that attained by semi-analytic models \citep{EAGLE}, and the observed atomic and molecular hydrogen content of galaxies out to $z\sim3$ \citep{EAGLE-HI,EAGLE-H2}.

\item{High resolution, $\sim 10$s of parsecs.} The {\bf advantages} are that it is possible to compare simulation outputs in detail with high-redshift galaxy images and spectra to discover the drivers of morphological changes as galaxies evolve, but the {\bf disadvantage} is that such simulations are so expensive computationally that it is as yet impossible to run enough cases to create statistical samples.  Leading examples of this approach are FIRE simulations led by Phil Hopkins \citep[e.g.,][]{FIRE} and the ART simulation suite led by Avishai Dekel and myself \citep[e.g.,][]{ZolotovART}.  We try to compensate for the small number of high-resolution simulations by using simulation outputs to tune semi-analytic models, which in turn use cosmological dark-matter-only simulations like Bolshoi to follow the evolution of $\sim10^6$ galaxies in their cosmological context \citep[e.g.,][]{Porter14a,Porter14b,Brennan15}.  
\end{itemize}

The high-resolution FIRE simulations, based on the GIZMO smooth particle hydrodynamics code \citep{GIZMO} with supernova and stellar feedback, including radiative feedback (RF) pressure from massive stars, treated with zero adjusted parameters, reproduce the observed relation between stellar and halo mass up to $M_{\rm halo}\sim10^{12} M_\odot$ and the observed star formation rates \cite{FIRE}.  FIRE simulations predict covering fractions of neutral hydrogen with column densities from $10^{17} {\rm cm}^{-2}$ (Lyman limit systems, LLS) to $> 10^{20.3} {\rm cm}^{-2}$ (DLAs) in agreement with observations at redshifts z=2-2.5 \citep{FIRE-HI}; this success is a consequence of the simulated galactic winds.  FIRE simulations also correctly predict the observed evolution of the decrease of metallicity with stellar mass \citep{FIRE-MZR}, and produce dwarf galaxies that appear to agree with observations \citep{FIRE-Dwarfs} as we will discuss in more detail below.


The high-resolution simulation suite based on the ART adaptive mesh refinement (AMR) approach 
\citep{kravtsovetal97,ceverino09} incorporates at the sub-grid level many of the physical processes relevant for galaxy formation.  Our initial group of 30 zoom-in simulations of galaxies in dark matter halos of mass $(1-30)\times10^{12} M_\odot$ at redshift $z=1$ were run at 35-70 pc maximum (physical) resolution \citep{Ceverino12,Ceverino15}.  The second group of 35 simulations (VELA01 to VELA35) with 17.5 to 35 pc resolution of halos of mass  $(2-20)\times10^{11} M_\odot$ at redshift $z=1$have now been run three times with varying inclusion of radiative pressure feedback (none, UV, UV+IR), as described in \citet{Ceverino14}.  RF pressure including the effects of stellar winds \citep{HopkinsQM12, FIRE} captures essential features of star formation in our simulations.  In particular, RF begins to affect the star-forming region as soon as massive stars form, long before the first supernovae occur, and the amount of energy released in RF greatly exceeds that released by supernovae \citep{Ceverino14,Trujillo-Gomez15}.  In addition to radiation pressure, the local UV flux from young star clusters also affects the cooling and heating processes in star-forming regions through photoheating and photoionization.  We use our \mcrx\ code \citep{jonsson06, jonsson+06,jonsson+10, JonssonPrimack} to make realistic images and spectra of these simulated galaxies in many wavebands and at many times during their evolution, including the effects of stellar evolution and of dust scattering, absorption, and re-emission, to compare with the imaging and photometry from CANDELS\footnote{CANDELS, the Cosmic Assembly Near-infrared Deep Extragalactic Legacy Survey, was the largest-ever Hubble Space Telescope survey, see \url{http://candels.ucolick.org/}.} and other surveys -- see Fig. 5 for examples including the effect of CANDELization (reducing the resolution and adding noise) to allow direct comparison with HST images.

\begin{figure}[t!]
\begin {center}
\includegraphics[width=1.0\textwidth]{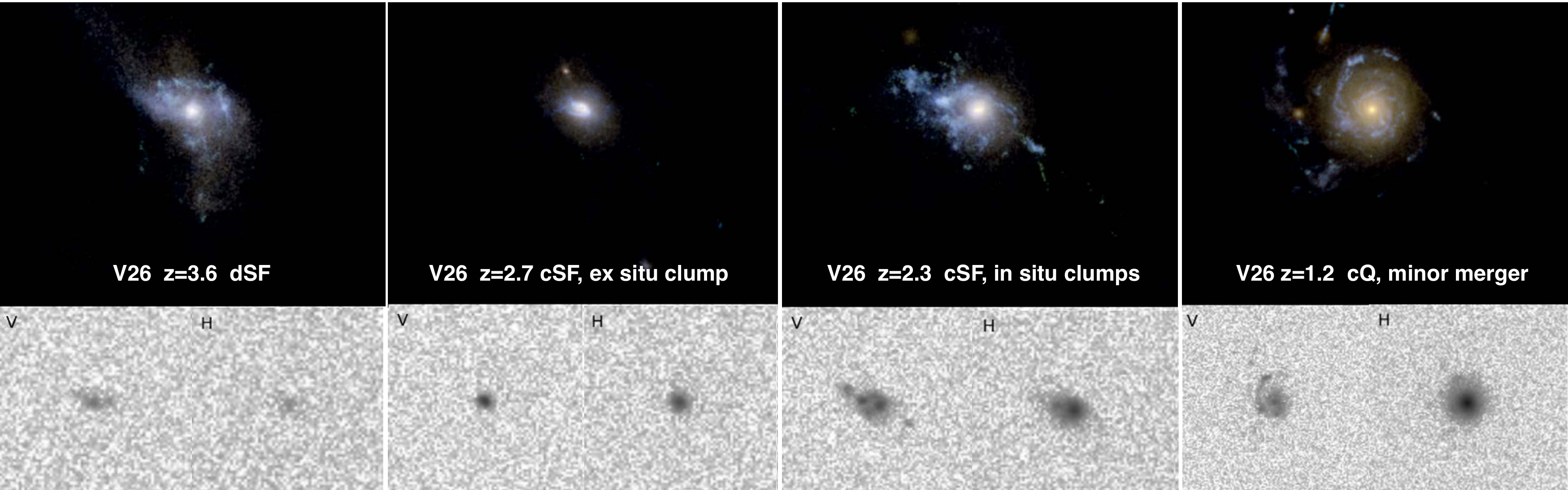}
\end {center}
\caption{Face-on images of Vela26 simulated galaxy with UV radiation pressure feedback, at four redshifts {\bf (a)} $z=3.6$ when it is diffuse and star forming (dSF); {\bf (b)} $z=2.7$ when it has become compact and star forming (cSF) with a red ex-situ clump; {\bf (c)} $z=2.3$ still cSF, now with in situ clumps apparent in the V-band image; {\bf (d)} compact and quenched (cQ) during a minor merger, with tidal features visible in the V-band image. Top panels: three-color composite images at high resolution; bottom panels:  CANDELized V and H band images.  The observed V band images correspond to ultraviolet radiation from massive young stars in the galaxy rest frame, while the observed H band images show optical light from the entire stellar population including old stars.  The CANDELS survey took advantage of the infrared capability of the Wide Field Camera 3, installed on the last service visit to HST in 2009.} 
\end{figure}

In comparing our simulations with HST observations, especially those from the CANDELS and 3D-HST surveys, we are finding that the simulations can help us interpret a variety of observed phenomena that we now realize are important in galaxy evolution.  One is the {\bf formation of compact galaxies}.  Analysis of CANDELS images suggested \citep{Barro13, Barro14a,Barro14b} that diffuse star-forming galaxies become compact galaxies (Òblue nuggetsÓ) which subsequently quench (Òred nuggetsÓ).  We see very similar behavior in our VELA simulations with UV radiative feedback \citep[][see Figure 2]{ZolotovART}, and we have identified in our simulations several mechanisms that lead to compaction often followed by rapid quenching, including major gas-rich mergers, disk instabilities often triggered by minor mergers, and opposing gas flows into the central galaxy \citep{Danovich15}.  

Another aspect of galaxy formation seen in HST observations is {\bf massive star-forming clumps} \citep[][and references therein]{Guo12, WuytsClumps13}, which occur in a large fraction of star-forming galaxies at redshifts z = 1-3 \citep{Guo15}.  In our simulations there are two types of clumps.  Some are a stage of minor mergers -- we call those ex situ clumps.  A majority of the clumps  originate in situ from violent disk instabilities (VDI) in gas-rich galaxies \citep{Ceverino12,Moody,Mandelker}.  Some of these in situ clumps are associated with gas instabilities that help to create compact spheroids, and some form after the central spheroid and are associated with the formation of surrounding disks.  We find that there is not a clear separation between these processes, since minor mergers often trigger disk instabilities in our simulations \citep{ZolotovART}.  

Star-forming galaxies with stellar masses $M_\ast \lsim 3\times10^{9} M_\odot$ at $z > 1$ have recently been shown to have mostly {\bf elongated (prolate) stellar distributions} \citep{vdWel14} rather than disks or spheroids, based on their observed axis ratio distribution.  In our simulations this occurs because most dark matter halos are prolate especially at small radii \citep{Allgood06}, and the first stars form in these elongated inner halos; at lower redshifts, as the stars begin to dominate the dark matter, the galaxy centers become disky or spheroidal \citep{CPD}.  

Both the FIRE and ART simulation groups and many others are participating in the Assembling Galaxies of Resolved Anatomy (AGORA) collaboration \citep{AGORA} to run high-resolution simulations of the same initial conditions with halos of masses $10^{10}$, $10^{11}$, $10^{12}$, and $10^{13} M_\odot$ at $z=0$ with as much as possible the same astrophysical assumptions.  AGORA cosmological runs using different simulation codes will be systematically compared with each other using a common analysis toolkit and validated against observations to verify that the solutions are robust -- i.e., that the astrophysical assumptions are responsible for any success, rather than artifacts of particular implementations. The goals of the AGORA project are, broadly speaking, to raise the realism and predictive power of galaxy simulations and the understanding of the feedback processes that regulate galaxy ``metabolism."

It still remains to be seen whether the entire population of galaxies can be explained in the context of $\Lambda$CDM.  A concern regarding disk galaxies is whether the formation of bulges by both galaxy mergers and secular evolution will prevent the formation of as many pure disk galaxies as we see in the nearby universe \citep{KormendyFisher08}.  A concern regarding massive galaxies is whether theory can naturally account for the relatively large number of ultra-luminous infrared galaxies.  The bright sub-millimeter galaxies were the greatest discrepancy between our semi-analytic model predictions compared with observations out to high redshift \citep{Somerville+12}.  This could possibly be explained by a top-heavy stellar initial mass function, or perhaps more plausibly by more realistic simulations including self-consistent treatment of dust \citep{Hayward,Hayward13}.  Clearly, there is much still to be done, both observationally and theoretically.
It is possible that all the potential discrepancies between $\Lambda$CDM and observations of relatively massive galaxies will be resolved by better understanding of the complex astrophysics of their formation and evolution.  But small galaxies might provide more stringent tests of 
$\Lambda$CDM.

\section{Smaller Scale Issues: Cusps}

Cusps were perhaps the first potential discrepancy pointed out between the dark matter halos predicted by CDM and the observations of small galaxies that appeared to be dominated by dark matter nearly to their centers \citep{FloresPrimack,Moore}.  Pure dark matter simulations predicted that the central density of dark matter halos behaves roughly as $\rho \sim r^{-1}$.  As mentioned above, dark matter halos have a density distribution that can be roughly approximated as $\rho_{\rm NFW} = 4\rho_s x^{-1}(1+x)^{-2}$, where $x\equiv r/r_s$ \citep{NFW}.  But this predicted $r^{-1}$ central cusp in the dark matter distribution seemed inconsistent with published observations of the rotation velocity of neutral hydrogen as a function of radius.

\begin{figure}
\includegraphics[width=\linewidth]{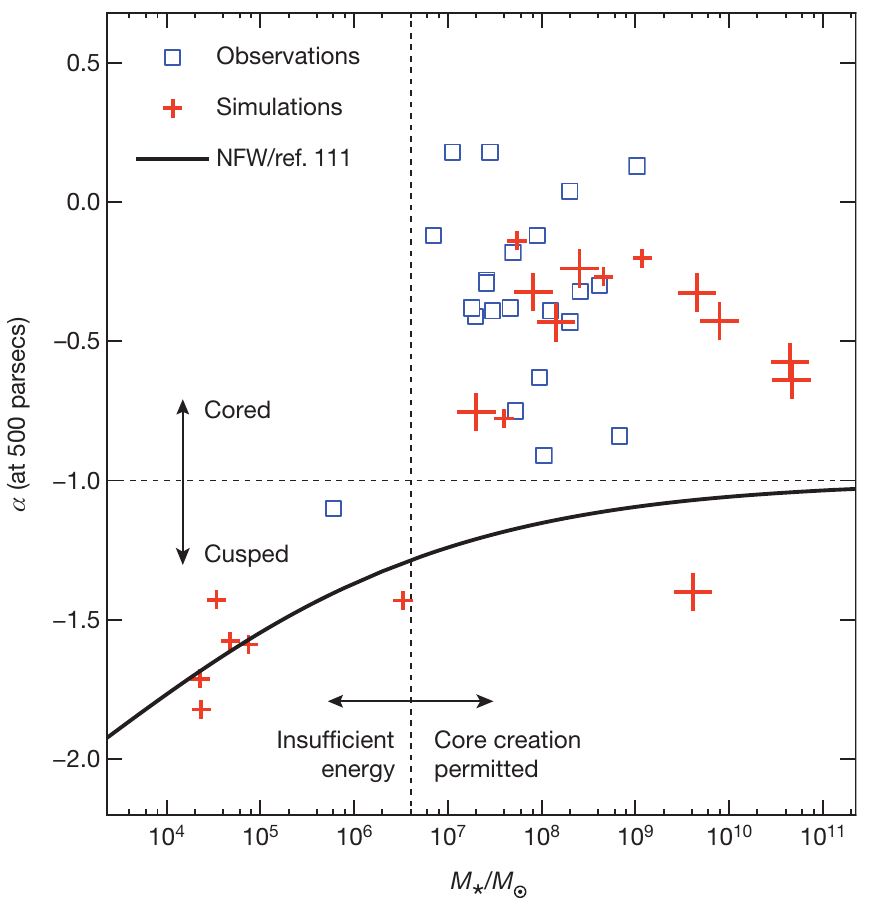}
\caption{Dark matter cores are generated by baryonic effects in galaxies with sufficient stellar mass.  The slope $\alpha$ of the dark matter central density profile $r^\alpha$ is plotted vs. stellar mass measured at 500 parsecs from simulations described in \citet{PontzenGovernato}. The solid NFW curve assumes the halo concentrations given by \citep{Maccio07}.  Large crosses: halos with $>5\times 10^5$ dark matter particles; small crosses: $> 5\times10^4$ particles. Squares represent galaxies observed by The HI Nearby Galaxy Survey (THINGS).  (Fig.~3 in Pontzen and Governato 2014.)}
\label{fig:6}
\end{figure}

In small galaxies with significant stellar populations, simulations show that central starbursts can naturally produce relatively flat density profiles \citep{Governato10,Governato12,PontzenGovernato,Teyssier13,Brooks14,BrooksZolotov,MadauShenGovernato14,
FIRE-Dwarfs,NipotiBinney}.  Gas cools into the galaxy center and becomes gravitationally dominant, adiabatically pulling in some of the dark matter \citep{BFFP86,Gnedin11}.  But then the gas is driven out very rapidly by supernovae and the entire central region expands, with the density correspondingly dropping.  Several such episodes can occur, producing a more or less constant central density consistent with observations, as shown in Fig.~6. 
The figure shows that galaxies in the THINGS sample are consistent with $\Lambda$CDM hydrodynamic simulations.  But simulated galaxies with stellar mass less than about $3\times10^6 M_\odot$ may have cusps, although \citet{FIRE-Dwarfs} found that stellar effects can soften the cusp in even lower-mass galaxies if the star formation is extended in time.  The observational situation is unclear.  In Sculptor and Fornax, the brightest dwarf spheroidal satellite galaxies of the Milky Way, stellar motions may imply a flatter central dark matter radial profile than $\rho \sim r^{-1}$\citep{WalkerPenarrubia11,AmoriscoEvans12,JardelGebhardt}.  However, other papers have questioned this \citep{JardelGebhardt13,BreddelsHelmi13,BreddelsHelmi14,RichardsonFairbairn}.

Will baryonic effects explain the radial density distributions in larger low surface brightness (LSB) galaxies?  These are among the most common galaxies.  They have a range of masses but many have fairly large rotation velocities indicating fairly deep potential wells, and many of them may not have enough stars for the scenario just described to explain the observed rotation curves \citep{KuziodeNaraySpekkens11}.  Can we understand the observed distribution of the $\Delta_{1/2}$ measure of central density \citep{AlamBullock02} and the observed diversity of rotation curves \citep{Maccio12CoreCusp,Oman}?  This is a challenge for galaxy simulators.  

Some authors have proposed that warm dark matter (WDM), with initial velocities large enough to prevent formation of small dark matter halos, could solve some of these problems.  However, that does not appear to work: the systematics of galactic radial density profiles predicted by WDM do not at all match the observed ones 
\citep{KuziodeNaray+10,Maccio12ContraWDM,Maccio13ContraWDM}.  WDM that's warm enough to affect galaxy centers may not permit early enough galaxy formation to reionize the universe \citep{Governato15WDM}.
Yet another constraint on WDM is the evidence for a great deal of dark matter substructure in galaxy halos \citep{ZentnerBullock03}, discussed further below.

\section{Smaller Scale Issues: Satellite Galaxies}

As the top panel of Fig.~3 shows, $\Lambda$CDM predicts that there are many fairly massive subhalos within dark matter halos of galaxies like the Milky Way and the Andromeda galaxy, more than there are observed satellite galaxies \citep{Klypin99,Moore99}.  This is not obviously a problem for the theory since reionization, stellar feedback, and other phenomena are likely to suppress gas content and star formation in low-mass satellites.  As more faint satellite galaxies have been discovered, especially using multicolor information from SDSS observations, the discrepancy between the predicted and observed satellite population has been alleviated.  Many additional satellite galaxies are predicted to be discovered by deeper surveys \citep[e.g.,][]{BullockStewart10}, including those in the Southern Hemisphere seen by the Dark Energy Survey \citep{DES15} and eventually the Large Synoptic Survey Telescope (LSST).

However, a potential discrepancy between theory and observations is the ``too big to fail" (TBTF) problem \citep{TooBig1,TooBig2}.  The Via Lactea-II high-resolution dark-matter-only simulation of a Milky Way size halo \citep{VL1,VL2} and the six similar Aquarius simulations \citep{Aquarius} all have several subhalos that are too dense in their centers to host any observed Milky Way satellite galaxy.  The brightest observed dwarf spheroidal (dSph)satellites all have $12 \lsim V_{\rm max} \lsim 25$ km/s.  But
the Aquarius simulations predict at least 10 subhalos with $V_{\rm max} > 25$ km/s.  These halos are also among the most massive at early times, and thus are not expected to have had their star formation greatly suppressed by reionization.  They thus appear to be too big to fail to become observable satellites \citep{TooBig2}.  

The TBTF problem is closely related to the cusp-core issue, since TBTF is alleviated by any process that lowers the central density and thus the internal velocity of satellite galaxies.  Many of the papers finding that baryonic effects remove central cusps cited in the previous section are thus also arguments against TBTF.  A recent simulation of regions like the Local Group found the number, internal velocities, and distribution of the satellite galaxies to be very comparable with observations \citep{Sawala14}.

Perhaps there is additional physics beyond $\Lambda$CDM that comes into play on small scales.  One possibility that has been investigated is warm dark matter (WDM).  A simulation like Aquarius but with WDM has fewer high-$V_{\rm max}$ halos \citep{Lovell}.  But it is not clear that such WDM simulations with the lowest WDM particle mass  allowed by the Lyman alpha forest and other observations \citep{Viel13,Horiuchi14} will have enough substructure to account for the observed faint satellite galaxies \citep[e.g.,][]{Polisensky11}, and as already mentioned WDM does not appear to be consistent with observed systematics of small galaxies.    

Another possibility is that the dark matter particles interact with themselves much more strongly than they interact with ordinary matter \citep{SIDM}.  There are strong constraints on such self-interacting dark matter (SIDM) from colliding galaxy clusters \citep{Harvey15ClusterSIDM,Massey15}, and in hydrodynamic simulations of dwarf galaxies SIDM has similar central cusps to CDM \citep{BastidasFry15}.   SIDM can be velocity-dependent, at the cost of adding additional parameters, and if the self-interaction grows with an inverse power of velocity the effects can be strong in dwarf galaxies \citep{Elbert}.
An Aquarius-type simulation but with velocity-dependent SIDM produced subhalos with inner density structure that may be compatible with the bright dSph satellites of the Milky Way  \citep{Vogelsberger}.  Whether higher-resolution simulations of this type will turn out to be consistent with observations remains to be seen.

\section{Smaller Scale Issues: Dark Matter Halo Substructure}

The first strong indication of galaxy dark matter halo substructure was the flux ratio anomalies seen in quadruply imaged radio quasars (``radio quads") \citep{MetcalfMadau01,DalalKochanek02,MetcalfZhao02}.  Smooth mass models of lensing galaxies can easily explain the observed positions of the images, but the predictions of such models of the corresponding fluxes are frequently observed to be strongly violated.  Optical and   X-ray quasars have such small angular sizes that the observed optical and X-ray flux anomalies can be caused by stars (``microlensing"), which has allowed a measurement of the stellar mass along the lines of sight in lensing galaxies \citep{Pooley12}.  But because the quasar radio-emitting region is larger, the observed radio flux anomalies can only be caused by relatively massive objects, with masses of order $10^6$ to $10^8 M_\odot$ along the line of sight.  After some controversy regarding whether $\Lambda$CDM simulations predict enough dark matter substructure to account for the observations, the latest papers concur that the observations are consistent with standard theory, taking into account uncertainty in lens system ellipticity \citep{MetcalfAmara12} and intervening objects along the line of sight \citep{XuMao+12,Xu15}.  But this analysis is based on a relatively small number of observed systems (Table 2 of \citet{ChenKZ11} lists the 10 quads that have been observed in the radio or mid-IR), and further observational and theoretical work would be very helpful.

Another gravitational lensing indication of dark matter halo substructure consistent with $\Lambda$CDM simulations comes from detailed analysis of galaxy-galaxy lensing \citep{GGlens10,GGlens12,Vegetti14}, although much more such data will need to be analyzed to get strong constraints.  Other gravitational lensing observations including time delays can probe the structure of dark matter halos in new ways \citep{KeetonMoustakas09}.  \citet{Hezaveh13,Hezaveh14} show that dark matter substructure can be detected using spatially resolved spectroscopy of gravitationally lensed dusty galaxies observed with ALMA.  \citet{Nierenberg} demonstrates that subhalos can be detected using strongly lensed narrow-line quasar emission, as originally proposed by \citet{MoustakasMetcalf03}. 

The great thing about gravitational lensing is that it directly measures mass along the line of sight.  This can provide important information that is difficult to obtain in other ways.  For example, the absence of anomalous skewness in the distribution of high redshift Type 1a supernovae brightnesses compared with low redshift ones implies that massive compact halo objects (MACHOs) in the enormous mass range $10^{-2}$ to $10^{10} M_\odot$ cannot be the main constituent of dark matter in the universe \citep{MetcalfSilk07}. The low observed rate of gravitational microlensing of stars in the Large and Small Magellanic clouds by foreground compact objects implies that MACHOs in the mass range between $0.6\times10^{-7}$ and $15 M_\odot$ cannot be a significant fraction of the  dark matter in the halo of the Milky Way \citep{EROS}.  Gravitational microlensing could even detect free-floating planets down to $10^{-8} M_\odot$, just one percent of the mass of the earth \citep{Strigari12}.

A completely independent way of determining the amount of dark matter halo substructure is to look carefully at the structure of dynamically cold stellar streams.  Such streams come from the tidal disruption of small satellite galaxies or globular clusters. In numerical simulations, the streams suffer many tens of impacts from encounters with dark matter substructures of mass $10^5$ to $10^7 M_\odot$ during their lifetimes, which create fluctuations in the stream surface density on scales of a few degrees or less.  The observed streams contain just such fluctuations \citep{Yoon,Carlberg12,CarlbergGH,CarlbergG}, so they provide strong evidence that the predicted population of subhalos is present in the halos of galaxies like the Milky Way and M31.  Comparing additional observations of dynamically cold stellar streams with fully self-consistent simulations will give more detailed information about the substructure population.  The Gaia spacecraft's measurements of the positions and motions of vast numbers of Milky Way stars will be helpful in quantifying the nature of dark matter substructure \citep{NganCarlberg,FeldmannSpolyar}.

\section{Conclusions}

$\Lambda$CDM appears to be extremely successful in predicting the cosmic microwave background and large-scale structure, including the observed distribution of galaxies both nearby and at high redshift.  It has therefore become the standard cosmological framework within which to understand cosmological structure formation, and it continues to teach us about galaxy formation and evolution.  For example, I used to think that galaxies are pretty smooth, that they generally grow in size as they evolve, and that they are a combination of disks and spheroids.  But as I discussed in Section 3, HST observations combined with high-resolution hydrodynamic simulations are showing that most star-forming galaxies are very clumpy; that galaxies often undergo compaction, which reduces their radius and greatly increases their central density; and that most lower-mass galaxies are not spheroids or disks but are instead elongated when their centers are dominated by dark matter.

$\Lambda$CDM faces challenges on smaller scales.  Although starbursts can rapidly drive gas out of the central regions of galaxies and thereby reduce the central dark matter density, it remains to be seen whether this and/or other baryonic physics can explain the observed rotation curves of the entire population of dwarf and low surface brightness (LSB) galaxies.  If not, perhaps more complicated physics such as self-interacting dark matter may be needed. But standard
$\Lambda$CDM appears to be successful in predicting the dark matter halo substructure that is now observed via gravitational lensing and stellar streams, and any alternative theory must do at least as well.

\vskip 0.5cm

\noindent Acknowledgment: My research is supported by grants from NASA, and I am also grateful for access to NASA Advanced Supercomputing and to NERSC supercomputers.

\bibliography{Tenerife-Primack}
\bibliographystyle{cambridgeauthordate}

\end{document}